\documentclass[preprint, superscriptaddress,
amsmath,amssymb,aps,showkeys,showpacs,
final,secnumarabic, nofootinbib]{revtex4-2}

\usepackage{cmap}
\usepackage[T1,T2A]{fontenc}
\usepackage[utf8]{inputenc}
\usepackage[russian,english]{babel}
\usepackage{color}
\usepackage{graphicx}
\usepackage{dcolumn}
\usepackage{bm}
\usepackage[unicode=true,colorlinks=true,linkcolor=magenta, urlcolor=blue, citecolor = blue,breaklinks]{hyperref}
\usepackage{multirow}
\usepackage{url}
\usepackage{breakurl}
\DeclareGraphicsExtensions{.eps}

\begin{document}

\title{Image Data Augmentation for the TAIGA-IACT Experiment\\with Conditional Generative Adversarial Networks} 

\def\addressa{Lomonosov Moscow State University, Skobeltsyn Institute of Nuclear Physics. 1(2), Leninskie gory, GSP-1, Moscow, 119991, Russia}
\def\addressb{Research Institute of Applied Physics. 20 Gagarin Boul., Irkutsk, 664003, Russia}

\author{\firstname{Yu.Yu.}~\surname{Dubenskaya}}
\email[E-mail: ]{dubenskaya@theory.sinp.msu.ru}
\affiliation{\addressa}
\author{\firstname{A.P.}~\surname{Kryukov}}
\affiliation{\addressa}
\author{\firstname{E.O.}~\surname{Gres}}
\affiliation{\addressb}
\author{\firstname{S.P.}~\surname{Polyakov}}
\affiliation{\addressa}
\author{\firstname{E.B.}~\surname{Postnikov}}
\affiliation{\addressa}
\author{\firstname{P.A.}~\surname{Volchugov}}
\affiliation{\addressa}
\author{\firstname{A.A.}~\surname{Vlaskina}}
\affiliation{\addressa}
\author{\firstname{D.P.}~\surname{Zhurov}}
\affiliation{\addressb}

\begin{abstract}
Modern Imaging Atmospheric Cherenkov Telescopes (IACTs) generate a huge amount of data that must be classified automatically, ideally in real time. Currently, machine learning-based solutions are increasingly being used to solve classification problems. However, these classifiers require proper training data sets to work correctly. The problem with training neural networks on real IACT data is that these data need to be pre-labeled, whereas such labeling is difficult and its results are estimates. In addition, the distribution of incoming events is highly imbalanced. Firstly, there is an imbalance in the types of events, since the number of detected gamma quanta is significantly less than the number of protons. Secondly, the energy distribution of particles of the same type is also imbalanced, since high-energy particles are extremely rare. This imbalance results in poorly trained classifiers that, once trained, do not handle rare events correctly. Using only conventional Monte Carlo event simulation methods to solve this problem is possible, but extremely resource-intensive and time-consuming. To address this issue, we propose to perform data augmentation with artificially generated events of the desired type and energy using conditional generative adversarial networks (cGANs), distinguishing classes by energy values. In the paper, we describe a simple algorithm for generating balanced data sets using cGANs. Thus, the proposed neural network model produces both imbalanced data sets for physical analysis as well as balanced data sets suitable for training other neural networks. 
\end{abstract}

\keywords{machine learning, data augmentation, GAN, image generation, gamma astronomy, IACT \\[5pt]}

\maketitle

\section{Introduction}\label{intro}

The TAIGA experiment (Tunka Advanced Instrument for cosmic ray physics and Gamma Astronomy) \cite{taiga-about} is a complex system for ground-based gamma-ray astronomy which includes detectors of different types. The TAIGA-IACT detectors \cite{taiga-iact-about} are Imaging Atmospheric Cherenkov Telescopes (IACTs) used to track extensive air showers (EASs) initiated by high energy gamma rays or charged cosmic rays (mostly protons). An EAS detected by an IACT we will call an event, while the data recorded by the telescope's camera will be called an image of the event. According to the type of primary particle that initiated the EAS, there are two types of events: an event initiated by a gamma quantum (gamma event) and an event initiated by a proton (proton event). Determining the type of event (classification) is one of the most important tasks in processing IACT data. Like other Cherenkov telescopes \cite{magic-about,hess-about,cta-about,veritas-about}, the TAIGA-IACT detectors produce a huge amount of images that must be classified automatically, ideally in real time. Currently, solutions based on machine learning (ML) are increasingly being used to solve classification problems. An important point in this approach is that training neural network classifiers requires properly prepared and labeled training data sets. Such data sets can be created based on both real data and artificially generated (synthetic) data.

When training a classifier the major issue is the imbalanced learning problem \cite{imbalance-learning-problem}. Saying that a data set is imbalanced with respect to some parameter means that the distribution of this parameter is far from uniform. In classification terms, an imbalanced training set means that some events that the classifier learns to recognize are rare. Therefore, such events have little effect on the learning process, and, accordingly, the classifier will not be able to recognize them correctly. To correct the imbalance, an approach called data augmentation \cite{data-augmentation} can be used. Data augmentation is a set of techniques applied to a data set used to create new samples that are slightly different from the existing ones. The simplest examples of such techniques are rotations, translations, cropping and flipping of images. A more complex and very promising approach to data augmentation involves the use of ML models to generate new samples by learning from existing samples. In astrophysics, ML-based data augmentation has already been successfully applied to generate synthetic light curves from variable stars \cite{generate-stars-light-curves}, as well as to improve the exoplanet detection \cite{exoplanet-detection} and variable star classification accuracy \cite{variable-stars-classification}. In this paper, we also focus on ML-based data augmentation, aiming to apply it to IACT data.

For real IACT data, the problem of imbalanced training is acute. Firstly, in real IACT data there is an imbalance in the types of events, since the number of detected gamma quanta is significantly less than the number of protons. Secondly, the energy distribution for events of the same type is also imbalanced, since high-energy events are rare phenomena. This is why synthetic IACT data, whose distribution varies depending on the task and the image generation method, is often preferred for training neural networks.

A well-established method for generating synthetic images is using special software that performs realistic Monte-Carlo simulations of both the EAS evolution \cite{corsika-about} and the response of the IACT system \cite{iact-responce-about}. By generating artificial data in this way, one can obtain the required number of events of a given type and with a given energy, so there is no problem of imbalanced learning. Therefore, using such model data to train classifiers seems to be an acceptable option. However, the problem is that the computational models of the underlying physical processes are very resource intensive and time-consuming. For some analysis purposes, the detailed information provided by the Monte Carlo simulation is redundant, so less complex and more efficient generation methods, such as ML-based generative models can be used.
When generating images using ML, the resulting data sets can be either balanced or imbalanced, depending on the training procedure. For non-ML applications such as physical analysis, the distribution of the generated sample is required to be the same as the distribution of the real IACT data, i.e. non-uniform. Therefore, some machine learning solutions \cite{jd-dlcp22,jd-dlcp23} focus on reproducing the distribution of the real IACT data in a sample of artificially generated events. However, as mentioned above, using such imbalanced data to train ML classifiers is inefficient. One way to solve the issue is to train another one neural network to generate balanced sets, but this is hardly an optimal solution because one will always have to keep two networks for generating balanced and unbalanced sets. Another way is to train one common network, generate an excess number of images with it, and then form data sets with the required distributions by discarding those images that do not have the desired parameters. Although the ML generation is very fast, the disadvantage of this method is a decrease in the final generation speed. This is due to the fact that to perform the selection, all the generated images have to be processed to determine their parameters, which also takes time.
In this paper, we propose a solution to quickly generate both imbalanced and balanced data sets of IACT images using a single conditional generative adversarial network (cGAN) \cite{cgan-about}. We describe a method for training such a cGAN, as well as an algorithm for data augmentation using this same network to tackle the imbalance in energy for gamma events.
  
\section{\label{sec:iact-img}IACT images and its parameters}
\begin{figure}
\includegraphics[width=8cm]{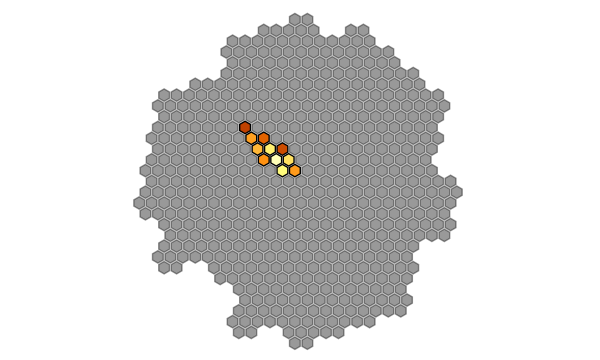}
\caption{\label{fig:gamma-image} Example of a hexagonal image recorded by the TAIGA-IACT detector.}
\end{figure}

The ground-based TAIGA-IACT detector records flashes of Cherenkov light using a camera that is a hexagonal array of about 600 photomultipliers (PMTs). Each PMT produces one image pixel, and all the pixels form a hexagonal image of the event (see Fig.~\ref{fig:gamma-image}).
The colored spot on each image is the region of the triggered PMTs. The colored spot has a more or less elliptical shape for both gamma events and proton events. However, primary particles of different types generate air showers that develop in different ways, so the gamma images are slightly different from the proton ones. In addition, both the energy value of the primary particle and the arrival direction affect the brightness, shape and location of the recorded ellipse. Thus, the parameters of the primary particle, such as its type, energy and arrival direction can be estimated from the appearance of the image. It should be noted that the estimation of these parameters is approximate, although it is correlated with the geometric parameters of the image.
The geometric parameters of the image, the so-called Hillas parameters \cite{hillas-about}, can be calculated exactly for each image. In this work we will mention the following two parameters: the image size parameter, which is the overall brightness of the image, which is the total sum of the values of all the triggered pixels, and the distance parameter, which is the distance between the centroid of the recorded ellipse and the center of the camera's field of view.

\section{\label{sec:proposed-approach}Generating both balanced and imbalanced data sets with a single \lowercase{c}GAN}
In general, the proposed approach to generate both balanced and imbalanced data sets with a single cGAN can be summarized as follows.
First, an imbalanced set of artificial Monte Carlo-simulated gamma images similar to real data recorded by TAIGA-IACT was used for training a cGAN. Using Monte Carlo simulated images for training has the advantage that the energy and other parameters of the corresponding events are exactly known. The training set was divided into classes artificially, based on the energy value of the primary particle that produced the EAS. The division was done in such a way that each class had the same number of images.
Once trained, a cGAN with such classes produces an imbalanced set of images that is close to the training set when asked to generate the same number of images of each class.
Data augmentation can be done by calculating the number of images for different classes that need to be generated in order for the final distribution to be uniform. That is, to get a balanced set of images using our cGAN the number of generated images of each class should be different and should depend of the class width. The algorithm for calculating the number of images for each class is given below in this paper in Section ~\ref{sec:augm-method}.

\section{\label{sec:gan-info} Preparing input data and training the \lowercase{c}GAN}

\subsection{\label{sec:training-set}The training set}
\begin{figure}
\includegraphics[width=8cm]{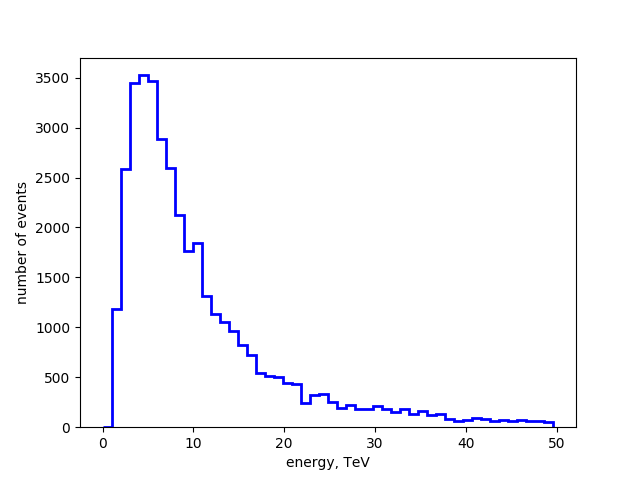}
\caption{\label{fig:energy-chart-in} The energy distribution of the training set.}
\end{figure}

For training we used a sample of images obtained using TAIGA Monte Carlo simulation software \cite{corsika-about, iact-responce-about}. The sample contains 40,000 images of gamma events with energies ranging from 0.5 to 50 TeV. This sample closely simulates the flow of real gamma events, which means that the energy distribution of this data set is extremely imbalanced (see Fig.~\ref{fig:energy-chart-in}).

\subsection{\label{sec:classes-div}Dividing IACT images into different classes}

Using the cGAN model implies a discrete approach, in which all images are divided into separate classes depending on the value of some parameter (or several parameters) of the image. In our previous work \cite{jd-dlcp22,jd-dlcp23} we divided images into classes based on the values of the image size Hillas parameter. This was done because the image size is easy to calculate for any image, and it is correlated with the energy of the primary particle. This was a good first approximation, but it is the energy of the primary particle that is the parameter of interest. That's why in this work, we trained our cGAN on images divided into classes based on the energy itself.

Just like the size value, the energy value is not limited to a discrete set of numbers, it can be any real number within a certain range. As can be seen in Fig.~\ref{fig:energy-chart-in}, the energy distribution has only one maximum, so we cannot distinguish several energy classes naturally, we can only do it artificially. In our previous work, we showed that the network learns better when each class contains the same number of images \cite{jd-dlcp22}. So, initially, we divided our training set of 40,000 images into 100 energy classes, containing 400 images in each class.
Our first results showed that, unlike size-based class dividing, energy-based class dividing results in incorrect cGAN learning, namely, the network does not recognize the difference between classes. This can be explained by the fact that, depending on the angle of arrival of the EAS, images with the same energy often look very different, and vice versa, images with different energies can sometimes look similar. To overcome this issue, when dividing into classes, we take into account not only the energy, but also the value of the distance Hillas parameter. The distance parameter was chosen because it is correlated with the direction of arrival of the EAS. So, in the training sample, we divided 10 classes by energy, and for each of these classes, we divided 10 classes by distance. In total, we got 100 classes with the same number of images.

\begin{figure}
\includegraphics[width=8cm]{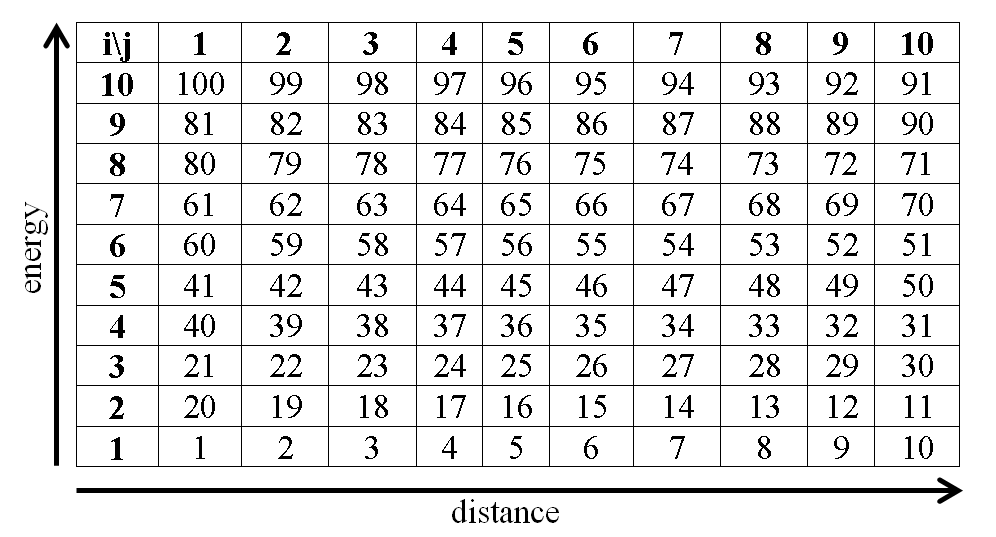}
\caption{\label{fig:class-numbers} Assigning class numbers in a snake-like pattern, $i$ - is an energy class number, $j$ - is a distance class number.}
\end{figure}

Our further research showed that with such an artificial division, for stable network training it is important that both parameters (energy and distance) change from class to class gradually, and not abruptly. To achieve this, we consider the set of classes as a two-dimensional array having 10 rows for energy and 10 columns for distance. To assign class numbers, we need to traverse the array in a snake-like pattern, as shown in Fig.~\ref{fig:class-numbers}. As can be seen from the figure, with this method of assigning class numbers for any two neighboring classes, both the energy and the distance do not differ much.

In more detail, the algorithm for assigning class numbers is as follows.
First, all images were sorted by energy in ascending order and divided into 10 classes so that each class had the same number of images (4000 images in our case). Thus, each energy class corresponds to a certain range of energies and has a serial number $i$ ($i$ from 1 to 10), where $i=1$ corresponds to the minimum energy range, and $i=10$ corresponds to the maximum one.
Then, for each class $i$, the following steps were performed.

All images of the energy class $i$ were sorted by distance in ascending order and divided into 10 classes so that each class had the same number of images (400 images in our case). Thus, each distance class corresponds to a certain range of distances and has a serial number $j$ ($j$ from 1 to 10), where $j=1$ corresponds to the minimum distance range, and $j=10$ corresponds to the maximum one. It should be noted that for different energy classes the distance ranges corresponding to the same $j$ numbers are generally different.
The final class numbers $k$ ($k$ from 1 to 100) are determined as follows:

\[ k=10(i-1) + j,~for~odd~i \]
\[ k=10(i-1) + (10 - j + 1) = 10i - j + 1,~for~even~i \]

Using energy and distance to divide the training set into classes, and assigning class numbers in a snake-like pattern significantly improved the results of network training compared to our first results.

\subsection{\label{sec:dataset-preproc}Training data set preprocessing}

\begin{figure}
\includegraphics[width=8cm]{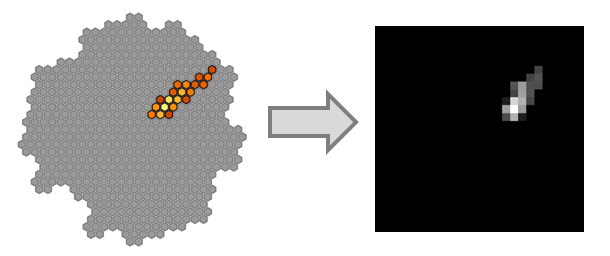}
\caption{\label{fig:img-preproc} A hexagonal image and the corresponding square image resulting from preprocessing.}
\end{figure}

It is worth noting that preprocessing of the training set is extremely important for generating images similar to those recorded by IACT. When preparing the training sample images, we applied cleaning, coordinate transformation, image resizing and pixel values normalization.
Normally, Monte Carlo images, as well as real data, contain noise from the night sky background fluctuations and electronic components. Image cleaning \cite{image-cleaning} is a conventional procedure to remove such noise thus leaving only images produced by a shower of secondary particles. In theory, cGANs can generate noisy images as well as cleaned ones. Our previous results \cite{jd-icrc21,jd-dlcp21,jd-grid21} show that our network reproduces cleaned images better. In order to teach our cGAN to generate cleaned images, we had to clean the images from the training set.
As noted earlier, the IACT images have a hexagonal structure. However, the current high performance cGAN implementations are designed for square images. That's why first, we generated rectangular images from hexagonal ones using axial coordinate transformation \cite{oblique-coordinates-about}. As a result, we got images of 31 by 30 pixels. Then, to make the images square, we resized each image to 32 by 32 pixels by adding two columns of zeros to the right and one row of zeros to the bottom.
Since the training set contains images with widely varying pixel values, we had to switch to a logarithmic scale by applying the logarithm function ln(1+x) to each pixel value of each image. Then, the pixel values were scaled to the range [-1, 1] to match the output of the generator model.
This is how we get a set of square grayscale images which we feed to the discriminator input during the training of our cGAN. An example of the original image and the image after preprocessing is shown in Fig.~\ref{fig:img-preproc}.

\subsection{\label{sec:cgan-arch}The proposed neural network model}

The proposed neural network model consists of three interacting components, each of which was trained separately: (1) a cGAN for image generation, 2) a neural network to determine the probability that an image is a gamma image (gamma likelihood) \cite{gamma-likelihood-about}, and 3) a neural network for energy reconstruction \cite{energy-reconstruction-about}. Neural networks 2 and 3 are results of the previous work and their architecture is beyond the scope of this paper. Here, we focus on the architecture of the cGAN network. The cGAN consists of a discriminator and a generator, and we'll describe the architecture of each of these networks separately.

\begin{figure}
\includegraphics[width=8cm]{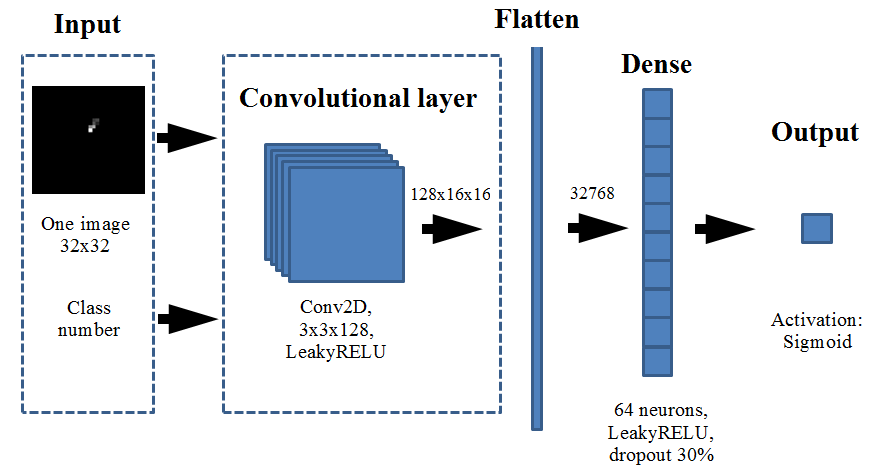}
\caption{\label{fig:discriminator} Architecture of the discriminator.}
\end{figure}

The architecture of the discriminator is shown in Fig.~\ref{fig:discriminator}. The discriminator is a convolutional neural network consisting of a convolutional layer with 3x3 filters followed by a dense (fully connected) layer with 64 neurons in it. The convolutional layers use a leaky ReLU function with alpha=0.2 as the activation function. The output layer uses sigmoid as the activation function.

\begin{figure*}
\includegraphics[width=16cm]{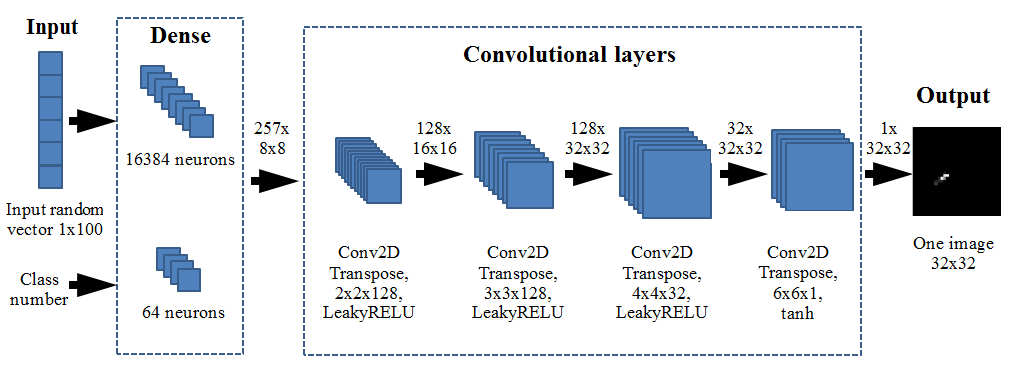}
\caption{\label{fig:generator} Architecture of the generator.}
\end{figure*}

The architecture of the generator is shown in Fig.~\ref{fig:generator}. The generator takes a random vector and a desired class number as input, and then uses transpose convolution to upsample until it gets an image of the required size. All layers use a leaky ReLU function with alpha=0.2 as the activation function. The output layer has one 6x6 filter and uses hyperbolic tangent as the activation function.

\subsection{\label{sec:cgan-training}cGAN training}

We used the above mentioned 100 artificial classes while training our cGAN. We have implemented the network with the proposed architecture using the TensorFlow \cite{tensorflow-about} software package. cGAN training at the GPU Tesla P100 with a batch size of 256 images and 500 epochs took about 6 hours. After training, generation of 1000 events of any class takes about 8 seconds. However, most of this time is spent loading libraries and the network model itself, so for comparison, generating 10,000 events takes 12 seconds, and generating 20,000 events takes 16 seconds.
The conventional Monte Carlo simulation software produces very accurate results, but is quite slow, generating an average of 1,000 images per hour. Using cGAN, therefore, speeds up the image generation process by several thousand times.

\subsection{\label{sec:training-results}Training results}

\begin{figure}
\includegraphics[width=8cm]{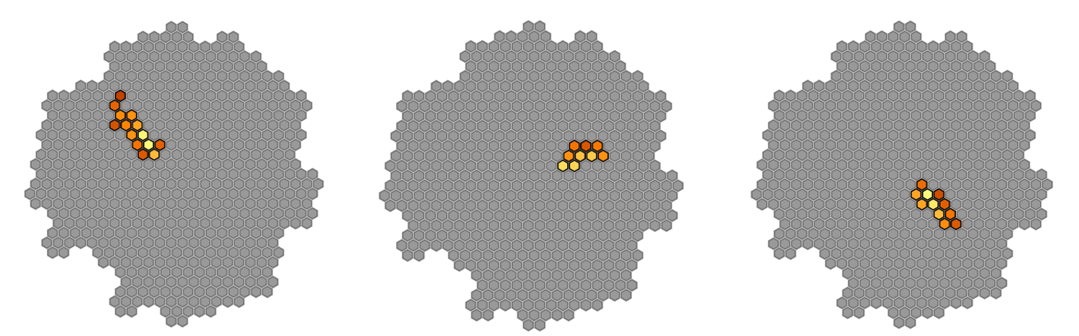}
\caption{\label{fig:generated-images} Examples of generated images.}
\end{figure}

The trained cGAN outputs square grayscale images with pixel values ranging from 0 to 1, exactly the same format as the training set. To get the image in the original format we first reverse-scale the images from the range [0, 1] using the previously stored maximum pixel value of the training set. We then convert the logarithmic pixel values to linear ones by applying an exponential function to each pixel value of each image: exp(x) - 1. Then we discard the extra pixels on the right and bottom of each image. Finally, the generated images are converted back to a hexagonal form.
The examples of the generated images are shown in Fig.~\ref{fig:generated-images}.

We checked the quality of the generated images using a separate neural network to determine the gamma likelihood. This check showed that over 98\% of the generated images are gamma images with a probability greater than 95\%, and less than 1\% of the generated images were misinterpreted as proton images. In further processing, images with low gamma likelihood are discarded.

\section{\label{sec:energy-real}Energy reconstruction for a sample of generated images}

\begin{figure}
\includegraphics[width=8cm]{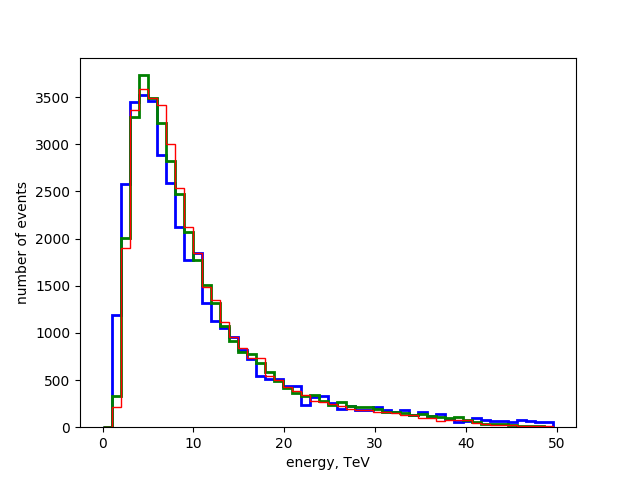}
\caption{\label{fig:energy-chart-real} Comparison of three imbalanced energy distributions: 1) the exactly known energy distribution of the training set (blue), 2) the energy estimate distribution for the training set (green), and 3) the energy estimate distribution for the generated sample (red).}
\end{figure}

To generate the imbalanced data set, we used the generator to create 400 images per each class, and this number was exactly the same as the number of images in each of the training set classes. So, we got 100 samples of 400 images and combined them into one larger sample, which gave us a total of 40,000 images. For this total sample, we used a separate neural network for energy reconstruction and then built a energy distribution for the sample. It should be noted, that the distance parameter is auxiliary, needed only for better network training, therefore further results are given only for the energy distribution.
To check the energy distribution for the generated sample, we compared three energy distributions: 1) the exactly known energy distribution of the training set, 2) the energy estimate distribution for the training set obtained by the neural network, and 3) the energy estimate distribution for the generated sample obtained by the neural network. The distribution 2 is used to additionally check the quality of energy reconstruction by the neural network. These three energy distribution are shown in Fig.~\ref{fig:energy-chart-real}.

When we compared the exactly known energy distribution for the training set and the reconstructed energy distribution for the cGAN-generated sample, we found that the chi-square test statistic is 1513 while the critical value corresponding for a 5\% significance level with 100 degrees of freedom is 124,34. However, comparing the reconstructed energy distribution for the training set and the reconstructed energy distribution for the cGAN-generated sample we found that the chi-square test statistic decreased to a value of 180. Although the chi-square test still shows that the difference is significant, the value of this criterion has decreased by a factor of eight, becoming much closer to the critical value. This result shows that the energy distribution of the generated sample is close enough to the energy distribution of the training set, but we need to further improve the quality of energy reconstruction.

\section{\label{sec:augm-method}Proposed method for data augmentation}

\begin{figure}
\includegraphics[width=8cm]{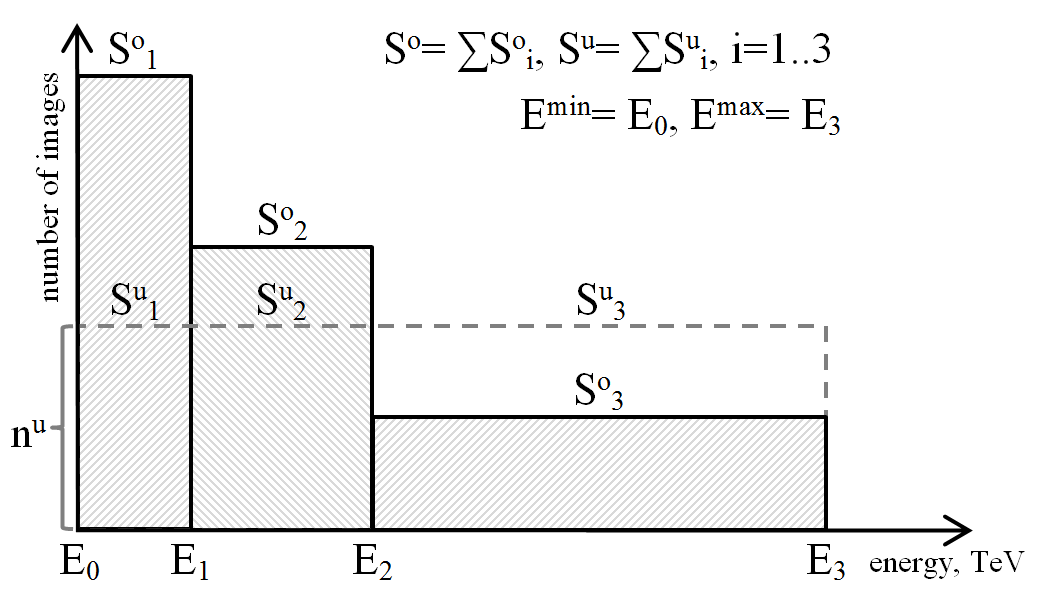}
\caption{\label{fig:augm-diagram} A diagram illustrating the augmentation procedure.}
\end{figure}

The task of augmentation is to generate a new sample so that the distribution by the parameter of interest is uniform. In this paper, we propose to generate a distribution close to uniform using a network trained on an unbalanced set, determining the number of images to be generated depending on the class width. For classes with wider bounds (such classes contain rare events), the number of images should be proportionally increased, respectively, for classes with narrower bounds, the number of images should be proportionally reduced.
Let us consider the simplest case, when images are divided into classes only using one parameter, for example, energy. A diagram illustrating the augmentation procedure is presented in Fig.~\ref{fig:augm-diagram}. In this example, the distribution is divided into 3 energy classes.

Let there be an original imbalanced energy distribution, divided into classes so that each class $i$ contains the same number of images. In Fig.~\ref{fig:augm-diagram}, the total number of images in class $i$ ($i=1..3$, where 3 is the total number of energy classes) is the area of the corresponding rectangle $S^o_i$. For the original distribution, all $S^o_i$ are equal to each other. Let us denote the total number of images in the original sample as $S^o$, $S^o=\sum S^o_i$.

For augmentation, we need to generate a new sample of $S^u$ images so that the energy distribution is uniform. Fig.~\ref{fig:augm-diagram} shows an example where $S^u=S^o$, but $S^u$ can take any value and does not depend on $S^o$. Let us denote as $S^u_i$ the total number of images of class $i$ that need to be generated, with $S^u=\sum S^u_i$. For a uniform distribution, the number of images per unit-length energy interval must be the same for all energy values. In Fig.~\ref{fig:augm-diagram}, the corresponding value is designated as $n^u$. It is easy to determine that $n^u= S^u/(E^{max}-E^{min})$, where $E^{min}$ and $E^{max}$ are, respectively, the minimum and maximum values of the energy. For the example shown in Fig.~\ref{fig:augm-diagram} $E^{min}=E_0$ and $E^{max}=E_3$. Accordingly, $S^u_i$ will be equal to the product of the value of $n^u$ and the class width $(E_i-E_{i-1})$, where $E_{i-1}$, $E_i$ are bounds of the $i$-th energy class:

\[ S^u_i = S^u (E_i-E_{i-1})/(E^{max}-E^{min}). \]

Let $K_i = (E_i-E_{i-1})/(E^{max}-E^{min})$, then $S^u_i = S^uK_i$.

Let us return to the more general case considered in this paper, where the division into classes is made using two parameters (energy and distance), and there are 10 energy classes with 10 distance classes in each. To determine the number of images in each class we introduce separate coefficients $K^E_i$ ($i=1..10$) for energy and $K^D_{i,j}$ ($i=1..10$, $j=1..10$) for distance, such that:

\[ K^E_i=(E_i-E_{i-1})/(E^{max}-E^{min}), \]
where $i=1..10$, $E_{i-1}$, $E_i$ are bounds of the $i$-th energy class, and $E^{min}$, $E^{max}$ are the minimum and maximum values of the energy.
\[ K^D_{i,j}=(D_{i,j}-D_{i,j-1})/(D^{max}_i-D^{min}_i), \]
where $i=1..10$, $j=1..10$, $D_{i,j-1}$, $D_{i,j}$ are bounds of the $j$-th distance class for the $i$-th energy class, and $D^{min}_i$, $D^{max}_i$ are the minimum and maximum values of distance for the $i$-th energy class.

Then we calculate the number of images for each class by multiplying the total number of the images to be generated by the corresponding coefficients $K^E_i$ and $K^D_{i,j}$:
\[ S^u_{i,j} = S^uK^E_iK^D_{i,j}, i=1..10, j=1..10. \]
An additional advantage of the proposed approach is the fact that the coefficients depend only on the class bounds, and therefore for a given cGAN they remain constant and can be calculated only once, which further increases the savings in machine time.

\section{\label{sec:augm-results}Data augmentation results}

\begin{figure}
\includegraphics[width=8cm]{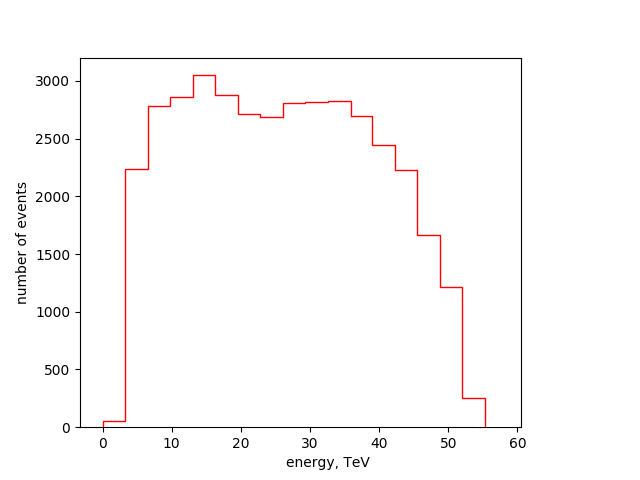 }
\caption{\label{fig:energy-chart-uni} The energy distribution for the sample of augmented data.}
\end{figure}

We calculated the coefficients $K^E_i$ and $K^D_{i,j}$ and the corresponding number of images for all our 100 classes as described in Section ~\ref{sec:augm-method}, then generated all of these images and reconstructed the energy for them. The energy distribution summed over all classes for this augmented data sample is shown in Fig.~\ref{fig:energy-chart-uni}.

As can be seen, the energy distribution is not uniform (the chi-square test statistic is about 4500), but is much closer to a uniform distribution than the imbalanced distribution shown in Fig.~\ref{fig:energy-chart-real}. The shape of this distribution can be explained as follows. Since the division into classes was made artificially, the energy distribution for a certain class of the generated sample is not uniform within the class bounds, but normal. Moreover, since the neural network performs not only interpolation, but also extrapolation of data, the distribution goes beyond the class bounds. When many classes have close bounds and the values for these classes are added together, the overall distribution looks uniform, as can be seen in Fig.~\ref{fig:energy-chart-uni}, in the range from 5 to 40 TeV.
However, the very high and very low energy classes have too wide bounds and in these energy ranges only they contribute to the total distribution. Because of this, for the energy range from 0.5 to 5 TeV, as well as from 40 to 50 TeV, the distribution differs significantly from uniform. In this work, we did not increase the number of classes because 400 images in one class is already quite a few. But if we increase the size of the training set and the number of energy classes, the class bounds will become closer, and the energy range in which the distribution is close to uniform will increase. But even now, if we take only the energy range from 5 to 40 TeV, where the distribution is close to uniform, we can already use the obtained results to train other neural networks.

\section{\label{sec:conclusion}Conclusion}

Simulating gamma images for the TAIGA-IACT experiment with a cGAN trained using energy-distance based class dividing yields promising results, as about 98\% of the generated gamma images are recognized as correct ones by third-party software with a probability of more than 95\%. At the same time, the rate of generation of such images using this cGAN is several thousand times higher than the rate of generation by the conventional Monte Carlo method.

The approach proposed in this paper allows for the rapid generation of both balanced and imbalanced data sets using a single network. Namely, we can generate images with either the original incoming energy distribution or perform data augmentation to obtain a nearly uniform energy distribution. In both cases, the corresponding energy distribution of the generated sample is relatively close to the desired energy distribution.

However, for both balanced and unbalanced data sets, the chi-square test shows that the difference in distributions is significant. On the other hand, validation of the training results of our network depends heavily on the methods for reconstructing the energy value of the generated images. For example, when we compared the exactly known energy distribution for the training set and the reconstructed energy distribution for the cGAN-generated sample, the chi-square test statistic was about 1500. But comparing the reconstructed energy distribution for the very same training set and the reconstructed energy distribution for the cGAN-generated sample we found that the chi-square test value statistic decreased to a value of 180. We expect that by improving the energy reconstruction, we will also obtain a closer distribution of the energy of the generated sample to the distribution of the training set. We also expect that increasing the number of training set images and, accordingly, increasing the total number of classes can also further improve the generation results, especially when generating a balanced data set.

Also it should be noted that there are other methods for handling class imbalance  \cite{class-imbalance-survey}. Some of them focus on altering the training data to decrease imbalance, while others involve modifying the learning or decision process to increase sensitivity towards the minority class, for example by taking a class penalty or weight into consideration. In our future work, we plan to investigate the applicability of these approaches to our problem.

\begin{acknowledgments}
The authors would like to thank the TAIGA collaboration for support and data provision. The work was carried out using equipment provided by the
MSU Development Program.
\end{acknowledgments}

\section*{FUNDING}
This study was supported by the Russian Science Foundation, grant no. 24-11-00136.

\end{document}